\documentstyle{article}
\begin{document}
\begin{flushright}
Preprint FIAN TD 10/94\\
To be published in JETP Lett.\\
October 1994\\
\end{flushright}

\begin{center}

\vspace{10 mm}

\Large\bf{MULTIPARTICLE PRODUCTION\\
AND STATISTICAL ANALOGIES}\\

\vspace{5 mm}

\large{I.M. DREMIN}

\vspace{3 mm}

\normalsize{Lebedev Physical Institute, Moscow, Russia;\\
E-mail address:   dremin@td.lpi.ac.ru }

\end{center}

\begin{abstract}
The study of singularities and zeros of the generating functions of
multiplicity distributions is advocated. Some hints from well known
probability distributions and experimental data are given. The statistical
mechanics analogies provoke to look for a signature of phase transitions.
The program of further experimental studies of the singularities is
formulated.
\end{abstract}

Multiplicity distributions in high energy collisions of various projectiles
and targets possess qualitatively similar (but quantitatively different!)
behaviour. That is why many fits by some well known probability distributions
have been tried. The ever more sensitive characteristics such as the ratio of
cumulant to factorial moments have been proposed \cite{d1} and have revealed
new features of experimental data \cite{d2}. Their understanding asks for
further experimental and theoretical studies. It is proposed here to pay more
attention to the structure of singularities and zeros of generating functions
of multiplicity distributions. It is especially appealing in view of possible
statistical analogies \cite{b3,h4}.

Let us define the generating function $G(z)$ of the probability distribution
$P_n$ by the relation
\begin{equation}
G(z) = \sum _{n=0}^{\infty }(1+z)^{n}P_n .
\end{equation}
In what follows, we often use also the function
\begin{equation}
\Phi (z) = \ln G(z) .
\end{equation}
The (normalized) factorial ($F_q$) and cumulant ($K_q$) moments of the
distribution $P_n$ are related to them by the formulae
\begin{equation}
G(z) = \sum _{q=0}^{\infty }\frac {z^q}{q!}\langle n\rangle ^{q}F_q ,
\;\; (F_{0}=F_{1}=1),
\end{equation}
\begin{equation}
\Phi (z) = \sum _{q=1}^{\infty }\frac {z^q}{q!}\langle n\rangle ^{q}K_q ,
\;\; (K_{1}=1) ,
\end{equation}
where $\langle n\rangle $ is the average multiplicity.

First we consider some distributions which provide analytical examples for the
nature of the singularities. We start with the fixed multiplicity (FM)
distribution when the sample of events of the same multiplicity ($n_0$) is
chosen, then proceed to Poisson distribution (P) as a reference to
independent emission processes and, finally, treat the gamma- ($\Gamma $),
negative binomial (NB) and lognormal (L) distributions widely used to fit
experimental data at high energies. The corresponding functions $\Phi (z)$
look like
\begin{equation}
\Phi ^{FM}(z) = n_{0}\ln (1+z) ,
\end{equation}
\begin{equation}
\Phi ^{P}(z) = z\langle n\rangle ,
\end{equation}
\begin{equation}
\Phi ^{\Gamma }(z) = -\mu \ln (1-\frac {\langle n\rangle }{\mu }\ln (1+z)) ,
\end{equation}
\begin{equation}
\Phi ^{NB}(z) = -k \ln (1-\frac {z\langle n\rangle }{k}) ,
\end{equation}
where $\mu $ and $k$ are the adjustable parameters. The lognormal distribution
is here the only one which is not determined by its moments. From the integral
representation of its generating function
\begin{equation}
\Phi ^{L}(z) \rightarrow -\ln \int _{0}^{\infty } \exp [-\frac {(\ln x -\nu
)^2}
{2\sigma ^2} + x\ln (1+z)]d(\ln x)
\end{equation}
it is easily seen that its convergence radius is given by the inequality
\begin{equation}
\vert z+1\vert _L \leq 1 ,
\end{equation}
i.e. the singularities come close to the point $z = 0$ but they are "soft" in
the sense that the normalization condition $G(0) = 1$ persists. For other
distributions the non-trivial (essential for our purposes) singularities are
situated at
\begin{equation}
z_{NB} = k/\langle n\rangle ,
\end{equation}
\begin{equation}
z _{\Gamma } = \exp (\mu /\langle n\rangle ) - 1 ,
\end{equation}
\begin{equation}
z_P = \infty ,
\end{equation}
\begin{equation}
z_{FM} = -1 .
\end{equation}
Let us note that NB and $\Gamma $-singularities are close to $z = 0$ if the
parameters $k$ and $\mu $ are much less than $\langle n\rangle $. It is
especially interesting because factorial and cumulant moments are calculated
as $q$-th derivatives of $G(z)$ and $\Phi (z)$ at that point and the nearby
singularity influences their behaviour substantially. In particular, it is
important for the ratio of the moments
\begin{equation}
H_q = K_{q}/F_{q} ,
\end{equation}
which is identically equal to zero for Poisson distribution, alternates sign
at each rank in case of fixed multiplicity, and is always positive for
$\Gamma $ and NB tending at asymptotically large ranks to zero as $q^{-k}$
\cite{d5}. The different type of behaviour is predicted in QCD with strong
decrease at low ranks followed by (quasi)oscillations at larger ranks
\cite{d1,d5,d6}. It would be interesting to guess what singularity governs
such a shape. There is no solution of the problem yet.

Let us get some guides from experimental data. In experiments with different
projectiles and targets the adjustable parameters are different and energy
dependent. Nevertheless, one can get qualitative estimate of the approximate
locations of the singular points. In $e^{+}e^{-}$-collisions, the NB-estimates
give rise to $k/\langle n\rangle \sim 1$ (see, e.g., \cite{a7}) and, therefore,
the singularity is situated at $z_{ee}\sim 1$, i.e. rather far from $z=0$. It
is much
closer to the origin in $hh$-collisions where (see, e.g., \cite{a8})
$k/\langle n\rangle \sim 10^{-1}$. The $AA$-data is not so definite \cite{c9}
(even though the lower statistics is slightly compensated by larger
multiplicity
) and give rise to $k/\langle n\rangle \leq 10^{-1}$ and, thus, to ever closer
(to the origin) singularity. The singularities move to the origin with energy
increase. Probably, these qualitative tendencies are related to somewhat
similar
regularities in the behaviour of the depth of the minimum of $H_q$ found
for various reactions (see \cite{d2,c9}) and to oscillations of $H_q$ at large
$q$ (see below). Moreover, the oscillations of experimental distributions about
the smooth NB-fit (see, e.g., \cite{a7}) could be connected with those
oscillations. Their physical meaning could correspond to various number of
subjets (ladders etc.) contributing at different multiplicities and should be
checked in Monte-Carlo models. Another possible source of oscillations due to
the cut-off of the multiplicity tail by conservation laws should die out
asymptotically \cite{d10}.

However, this cut-off plays an important role when one tries to restore the
generating function directly from experimental data. Actually, the series (1)
is replaced now by the polynomial in $z$
\begin{equation}
G_{N}(z) = \sum _{n=0}^{N} (1+z)^{n}P_n
\end{equation}
with $N$ equal to the highest observed multiplicity. Therefore $G_{N}(z)$ has
$N$ complex conjugate zeros
\begin{equation}
G_{N}(z) = \prod _{j=1}^{N} (1-\frac {z}{z_j}) .   \label{gn}
\end{equation}
It was shown by DeWolf \cite{d11} that the zeros cover a circle in the
complex $z$-plane for $ee$-events generated by JETSET Monte-Carlo program
at 1000 GeV. It reminds of Lee-Yang zeros \cite{y12} in statistical mechanics.
They do not seem always to close in onto the singularity of $G(z)$ on the real
axis when $N$ tends to infinity. It would be interesting to check the
interrelation of zeros of $G_{N}(z)$ and singularities of $G(z)$.

The cumulants are determined \cite{b3,d11} by the moments of zeros locations
\begin{equation}
K_{q} = -\frac {(q-1)!}{\langle n\rangle ^{q}}\sum _{j=1}^{N}z_{j}^{-q} =
-\frac {(q-1)!}{\langle n\rangle ^{q}}\sum _{j=1}^{N}\frac
{\cos q\theta _{j}}{r_{j}^{q}} ,
\end{equation}
where we denote $z_{j}=r_{j}\exp (i\theta _{j})$. Thus, the oscillations
mentioned above are related to the phases of zeros.

The study of singularities of the generating function becomes more fruitful if
one uses statistical mechanics analogies and recalls the Feynman fluid
model \cite {f13,b3,h4}. The generating function is analogous
to the partition function of the grand canonical ensemble and $\Phi (z)$
to free energy. The total rapidity range plays a role of the
volume and the variable $1+z$ is just the fugacity. One can define the
"pressure" $p(z)$ and the mean number of particles at given fugacity
$\langle n(z)\rangle $ (proportional to the usual pressure and density)
by the formulae
\begin{equation}
p(z) = \lim _{Y\rightarrow \infty }\frac {\Phi _{N}}{Y} ,    \label{pz}
\end{equation}
\begin{equation}
\langle n(z)\rangle _{N} = (1+z)\frac {\partial \Phi _{N}}{\partial z} ,
\label{nz}
\end{equation}
where $\Phi _{N}(z) = \ln G_{N}(z)$ and $\langle n(0)\rangle _{N} = \langle
n\rangle $. Let us note that the behaviour of $\langle n(z)\rangle _{N}$ in
the complex $z$-plane determined from experimental data should easily reveal
zeros $z_j$ of the function $G_N$ \cite{b3} since it has poles exactly at the
same loci $z_j$
\begin{equation}
\langle n(z)\rangle _{N} = \sum _{j=1}^{N} \frac {1+z}{z-z_{j}} . \label{np}
\end{equation}
The plots of $p(z)$ for experimental data about $ee$ and $hh$-reactions
extrapolated to $Y\rightarrow \infty $ have been shown in \cite{h4}. We
have checked that the latest LEP data (e.g., \cite{a7}) well coincide with
extrapolation used in \cite{h4} before the LEP data became available. The
authors of \cite{h4} claim that there is no phase transition in
$ee$-collisions.
The qualitative conclusion from Figs.3a and 3b of \cite{h4} is that $p(z)$
increases at $z>0$ much faster in non-diffractive $hh$-collisions as compared
to $ee$-collisions. It demonstrates that the $hh$-singularity is closer to
the origin that corresponds to above conclusions. The increase would be
even more drastic in case of $AA$-collisions (the data of EMU01 \cite{c9}
were used for estimates) but it is strongly influenced by single events with
very high multiplicity. Thus $AA$-analysis is hard to extend to large $z$.
Probably, it has a physical origin since $AA$-collisions are the most suspected
ones for phase transitions. Somewhat suspicious looks the constancy of $p(z)$
at $z<0$ for $hh$-collisions in Fig.3b of \cite{h4}. In statistical mechanics
it would be a signature for phase transition. If supported by further studies,
it would provide hints for theoretical speculations. Really, the problem of
phase transition in systems with relatively small number of particles should be
treated carefully. In particular, it depends on the steepness of increase of
$p(z)$ with $z$. Some criteria of it are awaited for. However, the similarities
may well happen to be mainly of formal nature and just the methods of analysis
are comparable. Nevertheless, some physical models based on the analogy have
been attempted \cite{s14,a15,c16,d17}.

Our preliminary qualitative results allow to formulate the further program of
analysis of experimental data which consists of determining
\begin{enumerate}
\item the radius of convergence of $G_N$ (1) according to Cauchi
($P_{n}^{1/n}$)
and D'Alambert ($P_{n}/P_{n-1}$) criteria,
\item the approach to the Carleman condition
$\sum _{n=1}^{\infty }F_{2n}^{-1/2n} = \infty $ at high energies
($N\rightarrow \infty $),
\item zeros of $G_{N}(z)$ (using the formulae (\ref{gn}) or (\ref{np})),
\item the behaviour of the "pressure" $p(z)$ (\ref{pz}),
\item the behaviour of the "multiplicity" $\langle n(z)\rangle $ (\ref{nz}),
\item the higher derivatives of $\Phi _N$ (the fractional derivatives can be
used also \cite{d18}, especially, in connection with the classification of the
phase transitions of non-integer order proposed recently \cite{h19}).
\end{enumerate}
The extrapolations to $Y\rightarrow \infty $ should be attempted. It is quite
probable that zeros locations will differ for different classes of processes
(diffractive and non-diffractive; two- and three-jets etc). The drastic change
in the behaviour of $\Phi _{N}$ or its derivatives must be carefully analysed
to look for a possible signature of the phase transition. In parallel,
the theoretical criteria of it in finite systems should be developed. We hope
that the first stage of the program formulated above can provide some new
insights into  the physics of multiparticle production. More detailed results
of it will be published elsewhere.

\vspace{2 mm}

\large{ACKNOWLEDGEMENTS}

I am indebted to E. DeWolf for pointing out the problem to me and sending
the unpublished paper \cite{d11}. The discussions with R. Hwa are acknowledged.
This work is supported by Russian fund for fundamental research (grant
94-02-3815) and by Soros fund (grant M5V000).


\begin{thebibliography}{20}

\bibitem{d1} I.M. Dremin, Phys. Lett. B313 (1993) 209.
\bibitem{d2} I.M. Dremin, V. Arena, G. Boca et al., Phys. Lett. B336 (1994)
119.
\bibitem{b3} K.J. Biebl, J. Wolf, Nucl. Phys. B44 (1972) 301.
\bibitem{h4} S. Hegyi, S. Krasznovszky, Phys. Lett. B251 (1990) 197.
\bibitem{d5} I.M. Dremin, R.C. Hwa, Phys. Rev. D49 (1994) 5805.
\bibitem{d6} I.M. Dremin, V.A. Nechitailo, JETP Lett. 58 (1993) 881.
\bibitem{a7} P.D.Acton et al. (OPAL), Zs. Phys. C53 (1992) 539.
\bibitem{a8} G.J. Alner et al. (UA5), Nucl. Phys. B291 (1987) 445.
\bibitem{c9} M.M. Chernyavskii, in Proc. 24th Int. Symposium on Multiparticle
Dynamics, Italy (1994), to be published by WSPC.
\bibitem{d10} I.M. Dremin, Physics-Uspekhi 164 (1994) 875.
\bibitem{d11} E.A. DeWolf, A note on multiplicity generating functions in the
complex plane (unpublished).
\bibitem{y12} C.N. Yang, T.D. Lee, Phys. Rev. D87 (1952) 404; 410.
\bibitem{f13} R.P. Feynman, Phys. Rev. Lett. 23 (1969) 1415.
\bibitem{s14} D.J. Scalapino, R.L. Sugar, Phys. Rev. D8 (1973) 2284.
\bibitem{a15} N.G. Antoniou, A.I. Karanikas, S.D.P. Vlassopulos,
Phys. Rev. D14 (1976) 3578; D29 (1984) 1470.
\bibitem{c16} P. Carruthers, I. Sarcevic, Phys. Lett. B189 (1987) 442.
\bibitem{d17} I.M. Dremin, M.T. Nazirov, Sov. J. Nucl. Phys. 55 (1992) 197;
2546.
\bibitem{d18} I.M. Dremin, JETP Lett. 59 (1994) 561.
\bibitem{h19} R. Hilfer, Phys. Rev. Lett. 68 (1992) 190; Phys. Rev. E48 (1993)
2466.

\end{thebibliography}
\end{document}